\documentclass[aps,prl,superscriptaddress,twocolumn,showpacs]{revtex4}
\usepackage{amsmath}
\usepackage{amssymb}
\usepackage{bm}
\usepackage{epsfig}
\usepackage{graphicx}
\usepackage{color}

\newcommand{\dar}{\downarrow}
\newcommand{\upar}{\uparrow}

\begin{document}

\title{Dynamics of spin-orbit coupled Bose-Einstein condensates in a random potential}
\author{Sh. Mardonov}
\affiliation{Department of Physical Chemistry, University of the Basque Country UPV/EHU,
48080 Bilbao, Spain}
\affiliation{The Samarkand Agriculture Institute, 140103 Samarkand, Uzbekistan}
\affiliation{The Samarkand State University, 140104 Samarkand, Uzbekistan}
\author{M. Modugno}
\affiliation{IKERBASQUE Basque Foundation for Science, Bilbao, Spain}
\affiliation{Department of Theoretical Physics and History of Science, University of the
Basque Country UPV/EHU, 48080 Bilbao, Spain}
\author{E. Ya. Sherman}
\affiliation{Department of Physical Chemistry, University of the Basque Country UPV/EHU,
48080 Bilbao, Spain}
\affiliation{IKERBASQUE Basque Foundation for Science, Bilbao, Spain}

\date{\today}

\begin{abstract}
Disorder plays a crucial role in spin dynamics in solids and condensed matter systems. 
We demonstrate that for a spin-orbit coupled Bose-Einstein condensate in a random
potential  two mechanisms of spin evolution, that can be characterized as ``precessional'' and ``anomalous'',  
are at work simultaneously.  The precessional mechanism, typical for solids, is due to the condensate displacement. 
The unconventional ``anomalous'' mechanism is due to the spin-dependent velocity producing the distribution
of the condensate spin polarization. The condensate expansion is accompanied 
by a random displacement and fragmentation, where it becomes 
sparse, as clearly revealed in the spin dynamics. Thus, different stages of the evolution 
can be characterized by looking at the condensate spin.

\end{abstract}

\pacs{03.75.Mn, 71.70.Ej, 03.75.Kk}

\maketitle

Spin-orbit coupling (SOC) is responsible for many fascinating properties of solids \cite{Dyakonov08}  and  
cold atoms \cite{zhaih2012,spielman2013}.  Bose-Einstein condensates (BECs) of (pseudo)spin-1/2 particles provide
novel opportunities for visualizing unconventional phenomena extensively studied 
experimentally (e.g. [\onlinecite{EXspielman2011,EXjin2012,EXqu2013,EXji2014}]) 
and theoretically (e.g. [\onlinecite{galitski2008,stringari2012,zhang2012,achilleos2013,ozawa2013,wilson2013,
lu2013,brandon2013,lindong2014}]).

In the presence of SOC, the spin of a particle rotates with a rate dependent on 
the particle's momentum. In disordered solids, randomization of momentum leads to the Dyakonov-Perel mechanism 
of spin relaxation \cite{Dyakonov} in macroscopic ensembles. 
{Studies of BEC spin dynamics in a random potential are strongly different from those in solids
in the following aspects: (i) one can access the evolution of a single wavepacket; (ii) one can study the 
effects of the anomalous spin-dependent velocity \cite{Adams} in different 
regimes of disorder and SOC, and (iii) the spin dynamics of a BEC is influenced by interatomic 
interactions inside each wavepacket, which are impossible for electrons.} 
Here we investigate these qualitatively new, unobservable in solids, effects in the spin evolution of
a {quasi one-dimensional} Bose-Einstein condensate. Usually, one is 
interested in the long time behavior, where localization takes over
\cite{Sanchez,Flach1,Pikovsky,Flach2,Larcher1,Aleiner,Larcher2,Min,Dujardin}. In the presence of SOC, the 
{localization} was studied in Ref.\cite{Zhou}. {Motivated by the fact that the evolving
spin density is well-defined for the experimental 
observation only at short time intervals, we consider here the initial stage of the evolution.} 

We consider a {SOC condensate  
tightly confined in the transverse directions to produce a quasi one-dimensional system,} 
subject to a random optical field producing a disorder potential $U_{\rm rnd}(x)$. 
The two-component wave function $\Psi(x,t)\equiv\left[\psi_{\upar}(x,t),\psi_{\dar}(x,t)\right]^{\rm T},$ {characterizing the spin 1/2 system  with
the density $|\Psi|^{2}=\left|\psi_{\upar}(x,t)\right|^{2}+\left|\psi_{\dar}(x,t)\right|^{2}$ 
normalized to the total number of particles $N\gg1,$} is obtained as a
solution of the nonlinear Schr\"{o}dinger equation 
$i\hbar\partial_{t}\Psi =\widehat{H}(t)\Psi.$ The effective Hamiltonian:
\begin{eqnarray}
\hspace{-0.9cm}\widehat{H}(t)&=&\frac{\widehat{p}^{2}}{2M}+\frac{M\omega^{2}(t)}{2}x^{2}+
U_{\rm rnd}(x)+{g}\hbar\omega_{0}a_{\rm ho}|\Psi|^{2} \nonumber \\
&+&\frac{\alpha}{\hbar}{\sigma}_{z}\widehat{p} +
\frac{\Delta_{Z}(t)}{2}\left({\bm \sigma}\cdot{\bm m}\right)\label{Hamiltonian1}
\end{eqnarray}
includes the interatomic interaction in the Gross-Pitaevskii form \cite{dalfovo1999,SPbook} with 
the dimensionless constant ${g}$ \cite{gcoupling}. 
Here $M$ is the particle mass and the frequency of the trap $\omega(t\le0)=\omega_{0}$, $\omega(t>0)=0$ 
corresponds to a sudden switch off. The energy quantum $\hbar\omega_{0}$ and the length
$a_{\rm ho}=\sqrt{\hbar/M\omega_{0}}$ represent the natural scales for the system description both for $t\le0$ 
and $t>0$. The SOC constant is $\alpha$ and ${\bm \sigma}$ is the Pauli matrix vector. 
The Zeeman term $\Delta_{Z}(t)\left({\bm \sigma}\cdot{\bm m}\right)/2,$ with ${\bm m}$ being the synthetic magnetic field direction,
prepares the initial spin state, and it is switched off at $t=0.$
 
\begin{figure}[t]
\begin{center}
\includegraphics[width=0.35\textwidth]{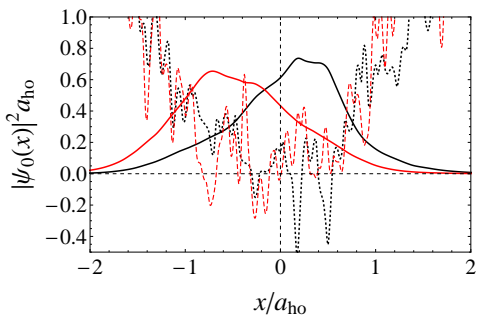}
\end{center}
\caption{(Color online.) Solid lines: two realizations of the ground state of a BEC with 
$g=0$, $\alpha=0$ in the total potentials (in $\hbar\omega_{0}-$units) shown by dashed lines. Here and below 
we use $U_{0}=\hbar\omega_{0}$, $\xi=a_{\rm ho}/32$, and $n\,=80/a_{\rm ho}$, corresponding 
to $\omega_{0}\tau\approx 4.$ The state with $\langle x(0)\rangle>0$ will be used
for calculations of the wavepacket evolution. 
For illustrative reasons, $U_{\rm rnd}(x)$ is rescaled by a factor $0.2$.}
\label{fig:FIG1}
\end{figure}

We assume a disorder produced by a random distribution of local potentials $U_{0}f(x-x_{j})$
at positions $x_{j}$ with the mean concentration $n$ as:
\begin{equation}
U_{\rm rnd}(x)=U_{0}\sum_{j}s_{j} f\left( x-x_{j}\right). 
\end{equation}%
Here $s_{j}=\pm 1$ is a random function of $j$ with $\langle s_{j}\rangle=0$ and $\langle U_{\rm rnd}(x)\rangle =0$. 
We consider $f\left(z\right)\equiv\exp
\left(-z^{2}/\xi^{2}\right)$ with a small width $\xi\ll a_{\rm ho}$ and assume a white-noise
distribution of impurities, resulting in \cite{Shklovskii}: 
\begin{equation}
\langle U_{\rm rnd}(x_{1})U_{\rm rnd}(x_{2})\rangle
=\langle U_{\rm rnd}^{2}\rangle e^{-{\left(x_{1}-x_{2}\right)^{2}}/{2\xi^{2}}},
\end{equation}%
with $\langle U_{\rm rnd}^{2}\rangle =\sqrt{{\pi}/{2}}U_{0}^{2}n\xi.$ 
{By using Fermi's golden rule we define} 
a scattering time $\tau={v_{\rm ho}\hbar^{2}}/{\pi n\,U_{0}^{2}\xi^{2}}$,
and a free path $\ell=v_{\rm ho}\tau$ 
with the velocity $v_{\rm ho}=\sqrt{\hbar\omega_{0}/M}$  [\onlinecite{oscillator}], 
and require $\ell/a_{\rm ho}\equiv\omega_{0}\tau\gg 1$ as a weak disorder condition of our interest.  
As the initial condition, we take 
the ground state of $\widehat{H}(t<0)$ in Eq. (\ref{Hamiltonian1}) at sufficiently strong $\Delta_{Z}>0$ and
${\bm m}=(-1,-1,0)/\sqrt{2}:$
\begin{equation}
\Psi(x,t=0)=\psi _{0}(x)\left[ 1,1\right] ^{\rm T}/\sqrt{2}.
\label{initial}
\end{equation}
Two possible $|\psi^{2}_{0}(x)|$ for $g=0$ are shown in Fig. \ref{fig:FIG1}. 

We characterize the spatial motion by three density-based quantities: the center of mass position
$\langle x(t)\rangle$ and two shape-related parameters such as 
the width $W(t)\equiv$ $\left[\langle x^2(t)\rangle-\langle x(t)\rangle^{2}\right]^{1/2}$ 
and the normalized participation ratio $\zeta(t)$ \cite{normalizedPR}: 
\begin{equation}
\zeta(t) =\left[\sqrt{2\pi }\int_{-\infty}^{\infty }\left\vert \Psi(x,t) \right\vert^{4}\frac{dx}{N^{2}}\right]^{-1}.
\end{equation}%

To study the dynamics, {we consider the force $F_{\Psi}$, 
defined as the derivative of the potential energy with respect to an infinitesimal ``virtual'' 
displacement $\delta x$}
as $\Psi(x,t)\rightarrow\Psi(x+\delta x,t).$ For a given realization of $U_{\rm rnd}(x)$ one obtains
\begin{equation}
F_{\Psi}=-\frac{1}{N}\int_{-\infty}^{\infty}U_{\rm rnd}(x)\partial_{x}\left|\Psi(x,t)\right|^{2}dx,
\label{FPsi}
\end{equation}
with the disorder-averaged \cite{Shklovskii} variance:
\begin{equation}
\langle F_{\Psi}^{2}\rangle=\frac{4\pi}{N^2}U_{0}^{2}n\,\xi^{2}\int\left[\partial_{x}\left|\Psi(x,t)\right|^2\right]^2{dx}.
\end{equation}%
This force with $\langle F_{\Psi}^{2}\rangle ^{1/2}\sim \sqrt{n\,}\xi\,U_{0}\zeta^{-3/2}(t)$, being 
determined by the spatial derivative of the density, is sensitive to the local structure of the condensate,  
resulting in a strong decrease for smooth density distributions. Since at $t\le0$ the ground state equilibrium 
requires $F_{\psi_{0}}=M\omega_{0}^{2}\langle x(0)\rangle$, 
the disorder potential causes the displacement of the wavepacket from the center 
of the trap (cf. Fig. \ref{fig:FIG1}) by 
$\langle x(0)\rangle=F_{\psi_{0}}/M\omega_{0}^{2}
\sim a_{\rm ho}\sqrt{a_{\rm ho}/\ell}.$

The spin state is fully described by the reduced density matrix ${\bm\rho}(t)$ with 
\begin{eqnarray}
&&\hspace{-10mm}\rho_{11}(t)=\int |\psi_{\upar}(x,t)|^{2}\frac{dx}{N},\quad \rho _{22}(t)=\int |\psi
_{\dar}(x,t)|^{2}\frac{dx}{N},\\
&&\hspace{-10mm}\rho_{12}(t)=\rho _{12}^{\ast }(t)=\int \psi _{\upar}^{\ast }(x,t)\psi _{\dar}(x,t)\frac{dx}{N}.
\label{denmat}
\end{eqnarray}%
The rescaled purity $P(t)$ $={2}{\rm tr}{\bm\rho}^{2}(t)-1$ with $0\le P(t)\le 1$ 
in the spin subspace is given by 
\begin{equation}
P(t)=1+4(|\rho_{12}(t)|^{2}-\rho_{11}(t)\rho_{22}(t)). 
\label{purity}
\end{equation}%
The spin components $\langle{\sigma}_{i}(t)\rangle \equiv
\mathrm{tr}\left({\sigma}_{i}{{\bm\rho}(t)}\right)$ yield $%
\rho _{12}(t)=\left(\langle{\sigma}_{x}(t)\rangle - i\langle 
{\sigma }_{y}(t)\rangle\right)/2$. With the initial state in Eq.(\ref{initial}) one obtains:
$\rho_{11}(t)=\rho_{22}(t)=1/2,$ $P(t)=\langle{\sigma}_{x}(t)\rangle^{2}+\langle{\sigma}_{y}(t)\rangle^{2}=$\, 
$4|\rho_{12}(t)|^{2},$ and $\langle{\sigma}_{z}(t)\rangle=0.$ 
The evolution of $\rho_{12}(t)$ is given by $\rho_{12}(t)=\left|\rho_{12}(t)\right|e^{i\phi_{12}(t)}$.
If $\left|\rho_{12}(t)\right|=1/2$ is conserved, the purity $P(t)=1,$ remains constant, and the spin evolution
is a precession with $\langle{\sigma}_{x}(t)\rangle=\cos\phi_{12}(t)$
and $\langle{\sigma}_{y}(t)\rangle=-\sin\phi_{12}(t)$. 
Equation (\ref{denmat}) shows that the spin dynamics 
is determined by the condensate structure 
in terms of the overlap of $\psi_{\upar}(x,t)$ and $\psi_{\dar}(x,t).$ 
This relation allows one to match the spin dynamics with the evolution of the wavepacket. 

To understand the effect of disorder on the spin evolution, one needs to consider
two mechanisms, which we will denote as
``precessional'' and ``anomalous''. To characterize the ``precessional'' mechanism,
we introduce the precession length $L_{\rm so}\equiv\hbar^{2}/M\alpha ,$ where 
$\phi_{12}(t)=2\left(\langle x(t)\rangle-\langle x(0)\rangle\right)/L_{\rm so}$ is due to the 
condensate displacement \cite{DP}. The ``anomalous''
mechanism \cite{Adams} appears 
owing to the fact that in the presence of SOC the
velocity operator becomes spin-dependent:
\begin{equation}
\widehat{v}=\frac{i}{\hbar}\left[ \frac{\widehat{p}^{2}}{2M}+\frac{\alpha}{\hbar}{\sigma }_{z}%
\widehat{p},\widehat{x}\right] =\frac{\widehat{p}}{M}+\frac{\alpha}{\hbar}{\sigma}_{z},  
\label{velocityspin}
\end{equation}
causing the experimentally observable spin-dipole oscillations \cite{EXjin2012} and a \textit{Zitterbewegung} \cite{EXqu2013}. 
As a result, any initial $\Psi(x,0)$, if it is not an eigenstate of ${\sigma}_{z}$, 
splits into spin-projected components. Therefore, the SOC leads, 
in addition to the precession, to a reduced $\left|\rho_{12}(t)\right|,$ 
decreasing the purity and modifying the spin evolution making it dependent on the spin density 
distribution inside the condensate.

\begin{figure}[t]
\begin{center}
\includegraphics[width=0.35\textwidth]{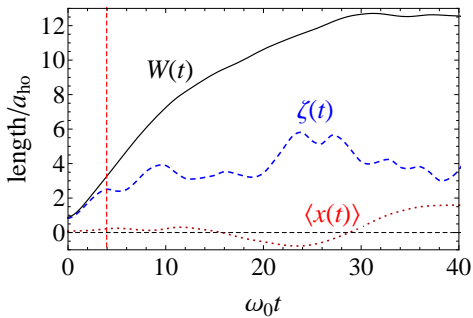}
\end{center}
\caption{(Color online) The wavepacket parameters (as  marked near the lines) of a noninteracting BEC. The picture shows 
the wavepacket fragmentation, where $W(t)$ exceeds $\zeta(t)$. The time dependences 
do not show qualitative changes with the realization 
of $U_{\rm rnd}(x)$. The vertical line corresponds to $t=\tau.$ }
\label{fig:FIG2}
\end{figure}

\begin{figure}[t]
\begin{center}
\includegraphics[width=0.35\textwidth]{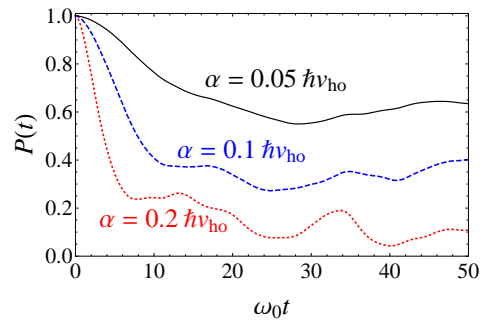}
\end{center}
\caption{(Color online) Purity of the spin state of a noninteracting BEC for different values of SOC. 
This Figure shows the crossover from decreasing $P(t)$ to the randomly oscillating behavior.}
\label{fig:FIG3}
\end{figure}

\begin{figure}[t]
\begin{center}
\includegraphics[width=0.35\textwidth]{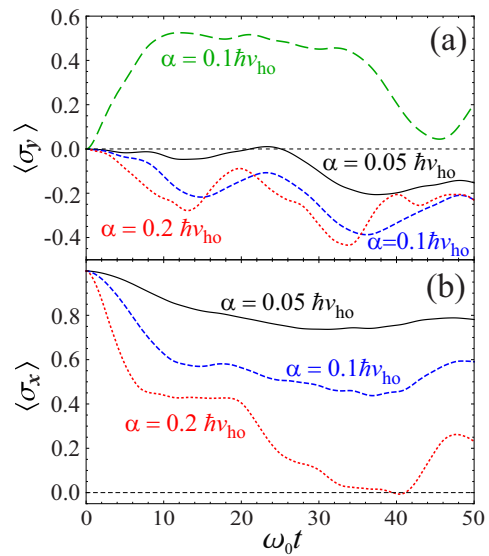}
\end{center}
\caption{(Color online) Time dependence of $\langle{\sigma}_{y}(t)\rangle$ and $\langle{\sigma}_{x}(t)\rangle$ 
for different values of $\alpha$.
The upper curve in (a) is for the realization 
of the random potential in Fig. \ref{fig:FIG1} with $\langle x(0)\rangle<0.$  
The behavior of $\langle\sigma_{y}(t)\rangle$ is mainly due to the spin precession.}
\label{fig:FIG4}
\end{figure}

{\it Noninteracting condensate.} In Fig. \ref{fig:FIG2} we present the
evolution of $W(t)$ and $\zeta(t)$  obtained from the numerical 
solution \cite{Tokatly}  of the Schr\"{o}dinger equation with $g=0.$ The corresponding $P(t)$ (Eq. (\ref{purity})) is
shown in Fig. \ref{fig:FIG3} and spin evolution is presented in Fig. \ref{fig:FIG4}. {A somewhat 
irregular behavior of the presented quantities corresponds to propagation of a wavepacket 
in a random potential with a finite correlation length.}

Immediately after releasing the harmonic potential, the condensate
starts moving due to the random force in the direction of $\langle
x(0)\rangle$ with the acceleration $F_{\psi_{0}}/M.$ The spin precesses accordingly 
to the $\langle x(t)\rangle-$displacement, 
with $\langle\sigma_{y}(t)\rangle=-\omega _{0}^{2}t^{2}\langle x(0)\rangle/L_{\rm so}.$  
The evolution of $\langle\sigma_{x}(t)\rangle=\sqrt{P(t)-\langle\sigma_{y}(t)\rangle ^{2}}$ has a different origin. 
In the precessional mechanism with
$P(t)=1$, one expects $\langle\sigma _{x}(t)\rangle=\sqrt{1-\langle\sigma_{y}(t)\rangle ^{2}},$
and, therefore, a quartic $\sim t^{4}$ initial behavior. 
However, the initial behavior in Fig. \ref{fig:FIG4}(a) is parabolic rather than 
quartic since the time dependence of $P(t)$ is important.
At small $t$ the wavefunction behaves as 
\begin{equation}
\Psi(x,t)=\left[\psi_{0}(x-\alpha t),\psi_{0}(x+\alpha t)\right]^{\rm T}, 
\label{psixt}
\end{equation}
and Eq.(\ref{denmat}) with $\Psi(x,t)$ in Eq. (\ref{psixt}) 
yields $P(t)=1-M\alpha^{2}E_{\rm kin}t^{2}/\hbar^{4},$ where $E_{\rm kin}\approx\hbar\omega_{0}/4$ is the initial kinetic energy.
In addition, the condensate starts to spread due to
the coordinate-momentum uncertainty. 
As a result of this spread, the force and the acceleration
decrease, and $\langle x(t)\rangle$ and $\phi_{12}(t)$ acquire a sub-$t^{2}$ dependence.
Thus, time-dependence of the spin is strongly related to the 
$U_{\rm rnd}(x)$ where the condensate moves. The behavior of $\langle
\sigma_{x}(t)\rangle $ at this stage is always due to the component separation,
demonstrating that two mechanisms of spin evolution are at work
simultaneously. On the time scale of the initial wavepacket broadening ($\sim\omega_{0}^{-1}$), the effect of the 
precession angle $1-\cos\phi_{12}(t)$ is of the order of $a_{\rm ho}^{2}/L_{\rm so}^{2}\times{a_{\rm ho}/\ell},$ while 
the change in the purity $1-P(t)$ is of the order of $a_{\rm ho}^{2}/L_{\rm so}^{2}.$ Therefore, for $\langle\sigma_{x}(t)\rangle,$
the effect of components separation is larger than the effect of precession by
a factor of $\sim\ell/a_{\rm ho}.$ 
Thus, at a weak disorder, the initial evolution of  
$\langle\sigma_{x}(t)\rangle$ is due to the purity decrease, while $\langle\sigma_{y}(t)\rangle$ evolves due to the 
spin precession. 

\begin{figure}[t]
\begin{center}
\includegraphics[width=0.35\textwidth]{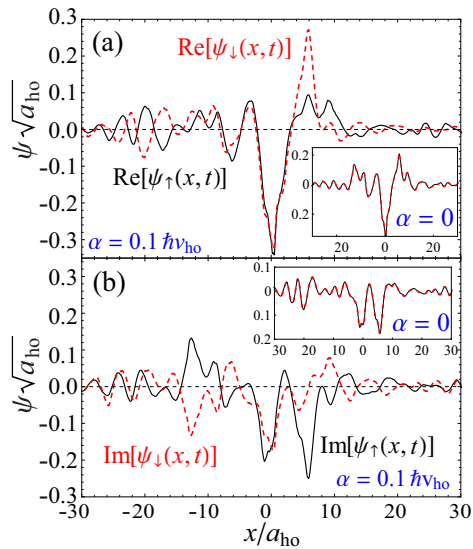}
\end{center}
\caption{(Color online) Components $\psi_{\upar}(x,t)$ and $\psi_{\dar}(x,t)$ (as marked near the plots)
at $t=18\omega_{0}^{-1}.$ 
The small integral overlap of these functions corresponds to the oscillating plateau in Fig. \ref{fig:FIG3}.  The 
insets show the two components at $\alpha=0$ \cite{Supplemental}.}
\label{fig:FIG5}
\end{figure}

At the following stage, for $t\ge\tau$, 
fragmentation of the condensate in the random potential begins,
and $\zeta(t)$ becomes smaller than $W(t)$ (Fig. \ref{fig:FIG2}).  
The density distribution becomes relatively sparse and consists of
several peaks of different width, in agreement with Ref. \cite{Flach1}. 
At this stage, 
the purity is related to the details of the wave function components. The evolution of  
$P(t)$ in Fig. \ref{fig:FIG3} shows a crossover to the oscillating plateau at time satisfying condition $2\alpha t/\hbar\sim\zeta(t).$ 
The reasons for the change in the purity (Fig. \ref{fig:FIG3}) and the corresponding spin dynamics 
(Fig. \ref{fig:FIG4}) can be understood from Fig. \ref{fig:FIG5},
showing ${\rm Re}[\psi_{\uparrow,\downarrow}(x,t)]$ and  
${\rm Im}[\psi_{\uparrow,\downarrow}(x,t)]$ after the fragmentation  
{in a random potential has produced the irregular shape of the wavefunction \cite{Supplemental}.} 
In the presence of SOC, the spinor components are considerably different,
and their relative oscillations lead to a decrease in $\left|\rho_{12}(t)\right|$, resulting in 
the purity decrease. If $2\alpha\tau/\hbar\le a_{\rm ho},$ the purity does not fall to zero and the
spin length $\sqrt{P(t)}$  remains approximately a constant at this stage. 
The oscillations correlate with the displacement 
$\langle x(t)\rangle$ changing on the time scale of the order of $\tau$ due to a nonvanishing force $F_{\Psi}.$
The spin precesses with $\langle\sigma_{y}(t)\rangle\sim \langle x(t)\rangle/L_{\rm so}$ 
as can be seen in Fig. \ref{fig:FIG4}(a). Since $\left|\langle x(t)\rangle\right|$ 
is of the order of $\zeta(t)\sim\ell\ll\,L_{\rm so},$
at sufficiently weak SOC $\langle\sigma_{y}(t)\rangle\ll 1.$

\begin{figure}[t]
\begin{center}
\includegraphics[width=0.35\textwidth]{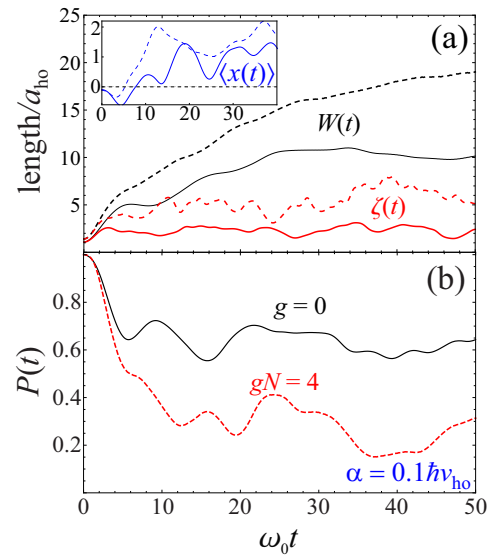}
\end{center}
\caption{(Color online) (a) The wavepacket parameters for $g=0$ (solid lines)
and for $gN=4$ (dashed lines); $\alpha=0.1\hbar\,v_{\rm ho}$.  (b) The spin purity for $\alpha=0.1\hbar\,v_{\rm ho}$. 
This Figure is obtained for a different realization of disorder with respect to that in Fig. \ref{fig:FIG2}.}
\label{fig:FIG6}
\end{figure}

{\it Effects of interaction.} Here we address the effect of a moderately strong repulsion ${g}N\sim 1$, 
causing an increase in $W(t)$ and 
$\zeta(t),$ as shown in Fig. \ref{fig:FIG6}, and, as a result, decreasing the random force (\ref{FPsi}) 
acting at the condensate. The repulsion accelerates the packet spread and the purity decreases faster due to the resulting
decrease in $\left|\rho_{12}(t)\right|$.  As a result, the effect of disorder in the spin dynamics weakens.  
The comparative behavior of purity is shown in Fig. \ref{fig:FIG6}.
The corresponding spin evolution is presented in \cite{Supplemental}. 
At $gN\gg 1,$ the spin dynamics becomes mainly interaction-determined.

\textit{Conclusions.} We have studied the dynamics of {a single wavepacket} of spin-orbit coupled 
Bose-Einstein condensate in a random potential on a  time scale of tens 
of the scattering times. We have predicted experimentally observable features 
of the spin evolution such as the dependence of the spin dynamics 
on the initial state and a qualitative difference between spin components including a crossover from 
decreasing to a plateau-like behavior. 
The {striking} feature of this process is that two different 
mechanisms of spin dynamics - a precessional one and the other one
due to the change in the spatial overlap of the spin
components, take place simultaneously. The former mechanism is due to the spin precession related to the 
condensate displacement, while the latter one, not seen in solids, is caused by the spatial separation of the spin 
components attributed to the anomalous spin-dependent velocity \cite{Adams}. 

\begin{acknowledgments} 
This work was supported by the University of Basque
Country UPV/EHU under program UFI 11/55, Spanish MEC (FIS2012-36673-C03-01
and FIS2012-36673-C03-03), and Grupos Consolidados UPV/EHU del Gobierno
Vasco (IT-472-10). S.M. acknowledges EU-funded Erasmus Mundus Action 2
eASTANA, \textquotedblleft evroAsian Starter for the Technical Academic
Programme\textquotedblright\ (Agreement No. 2001-2571/001-001-EMA2).
\end{acknowledgments} 


\newpage

\begin{widetext}

\begin{center}

\large{\textbf{Supplemental Material for ``Dynamics of spin-orbit coupled Bose-Einstein condensate in a random potential''}}

\end{center}

\section{Spin components of condensate wavefunctions.}

Here we present wavefunctions of spin-orbit coupled condensate at different stages, corresponding to the times before and after the fragmentation. 

\begin{figure}[h]
\begin{center}
\includegraphics[width=0.45\textwidth]{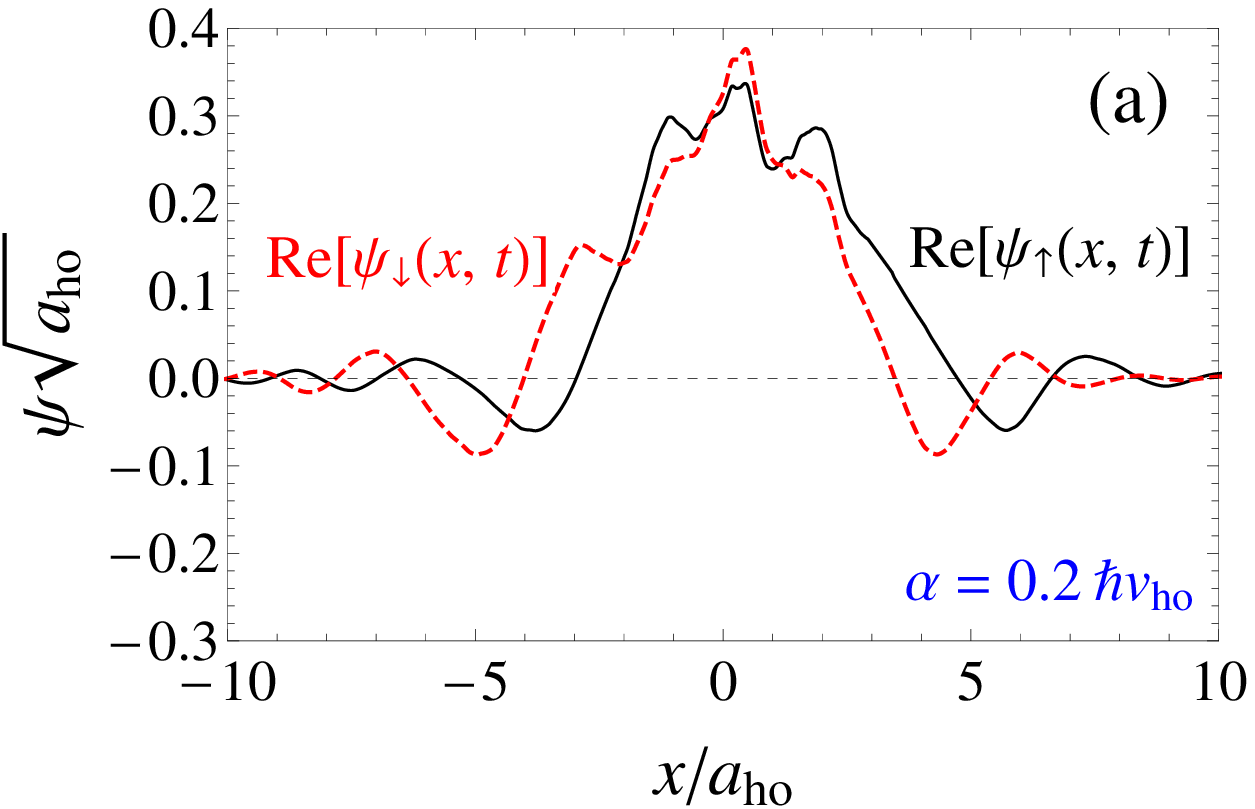} \vspace{0.5cm} \includegraphics[width=0.45\textwidth]{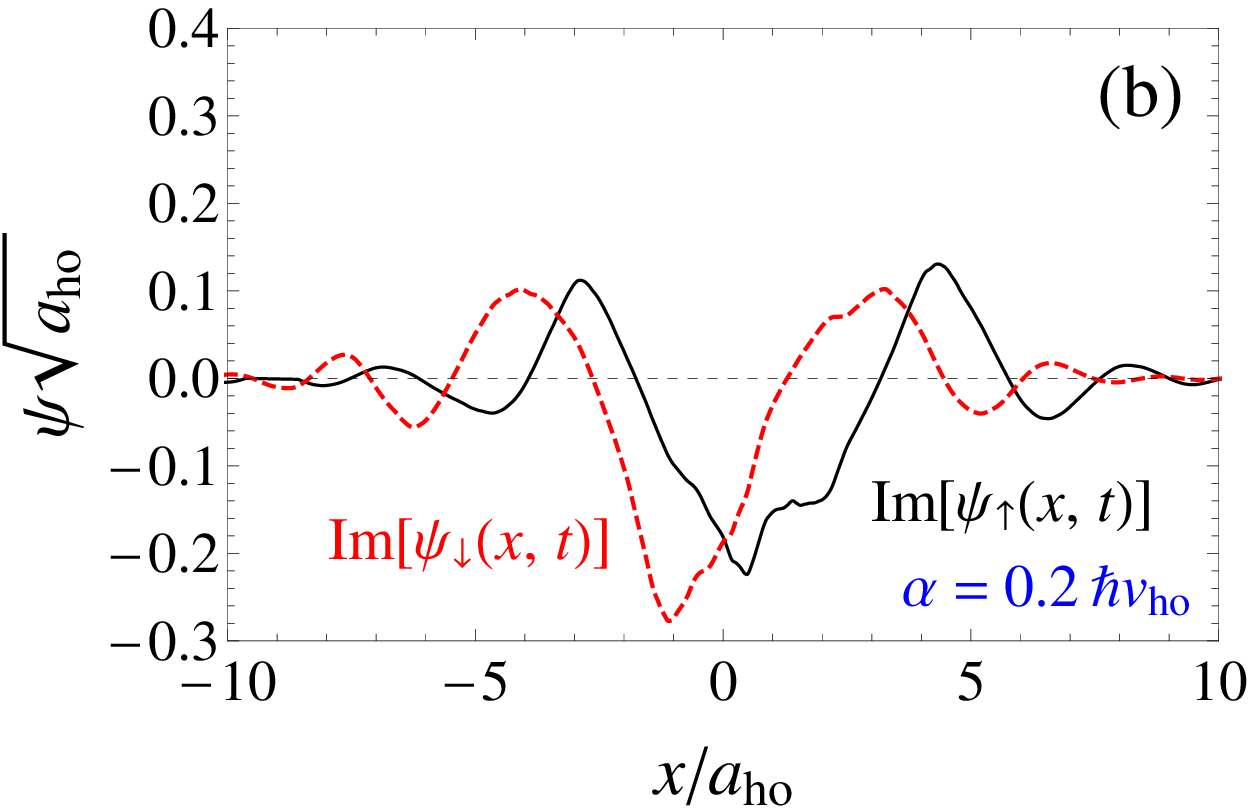}\\
\includegraphics[width=0.45\textwidth]{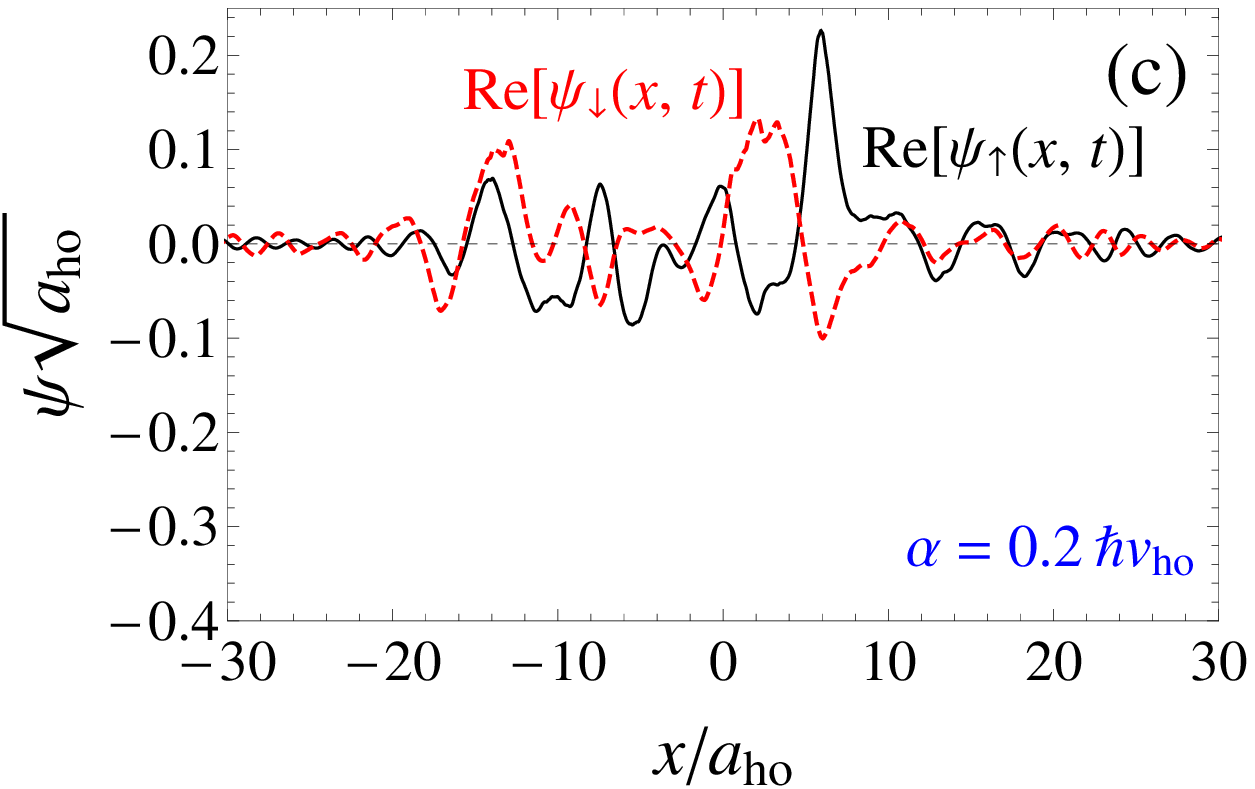} \vspace{0.5cm} \includegraphics[width=0.45\textwidth]{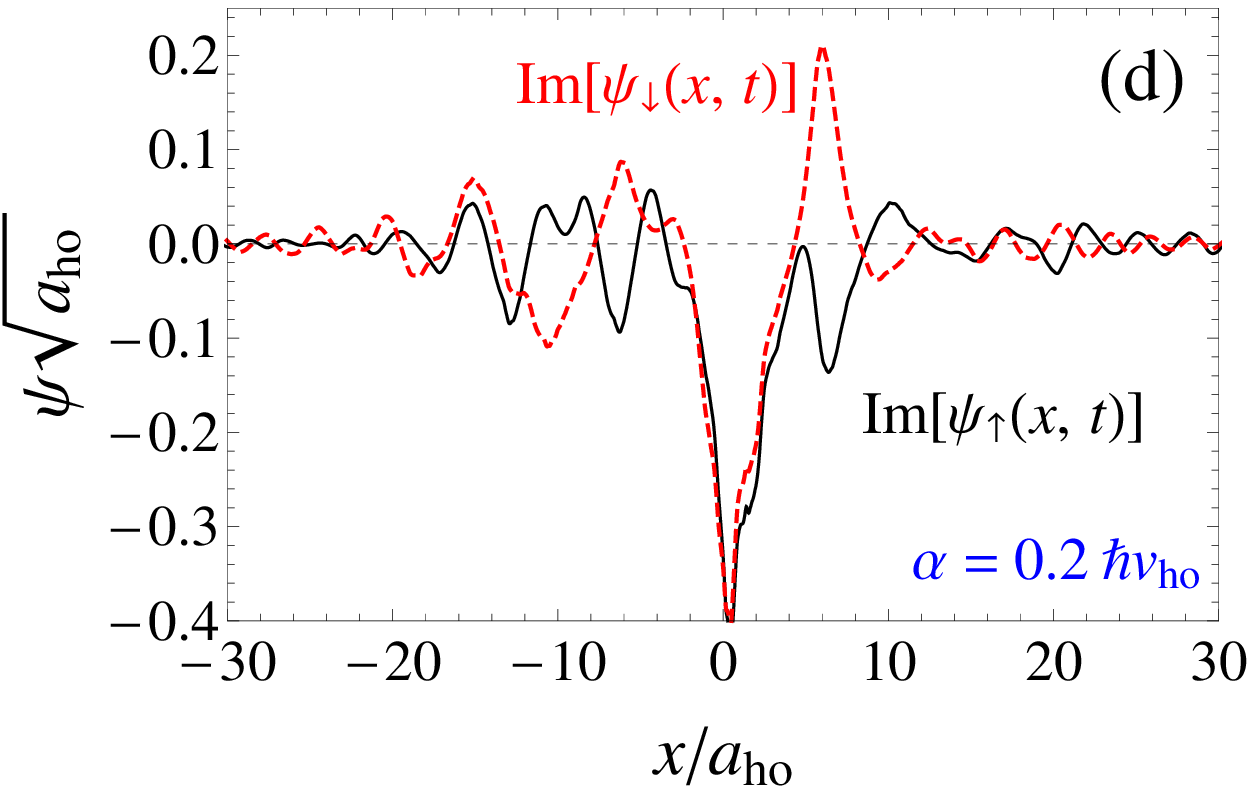}
\end{center}
\caption{Real and imaginary components of the wavefunction at different times for $\alpha=0.2{\hbar}v_{\rm ho}$. 
The upper row corresponds to a relatively 
short time $t=3\omega_{0}^{-1}$ where the condensate is not yet fragmented, while the lower row corresponds to the fragmented BEC
at $t=12\omega_{0}^{-1}$. Panels (a) and (b) show that before the fragmentation starts, the spin-projected components  
$\psi_{\uparrow}(x,t)$ and $\psi_{\downarrow}(x,t)$ 
are very similar. At the fragmentation a multiple peak structure is formed and the spinor components become 
considerably different to decrease the rescaled purity $P(t)$ of the system (Eq. (10) of the main text). The relative stabilization and oscillations of the  
purity are due to the fact that for the multiple peak structure there is no a monotonous decrease in $\left|\rho_{12}(t)\right|$ 
(as defined in Eq. (9) of the main text) with time.}
\label{fig:FIG6:1}
\end{figure}

\section{Effect of interatomic repulsion on spin dynamics.}

Here we present the effect of moderate repulsion ${g}N=4$ on the spin dynamics.

\begin{figure}[h]
\begin{center}
\includegraphics[width=0.45\textwidth]{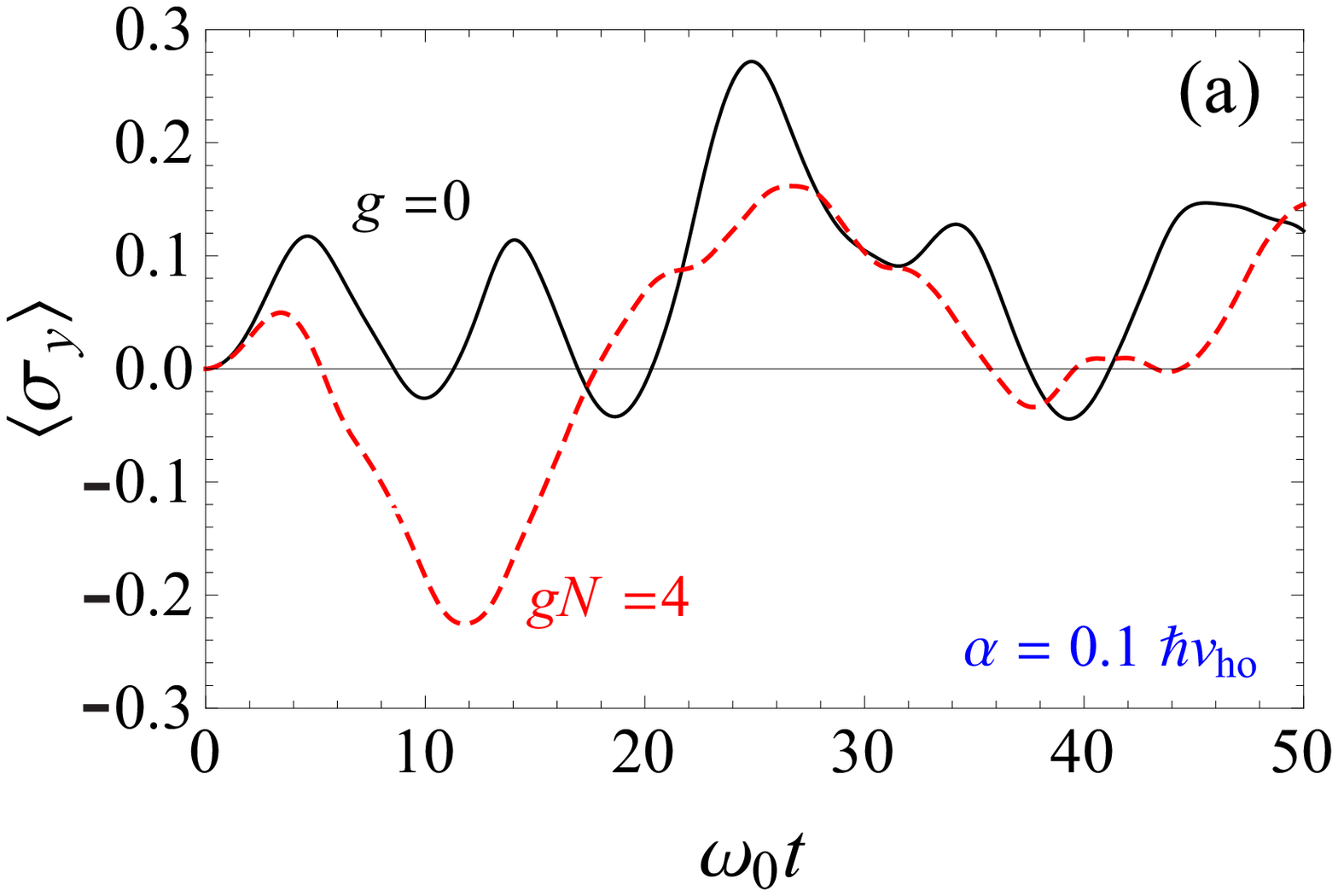}\vspace{0.5cm}\includegraphics[width=0.45\textwidth]{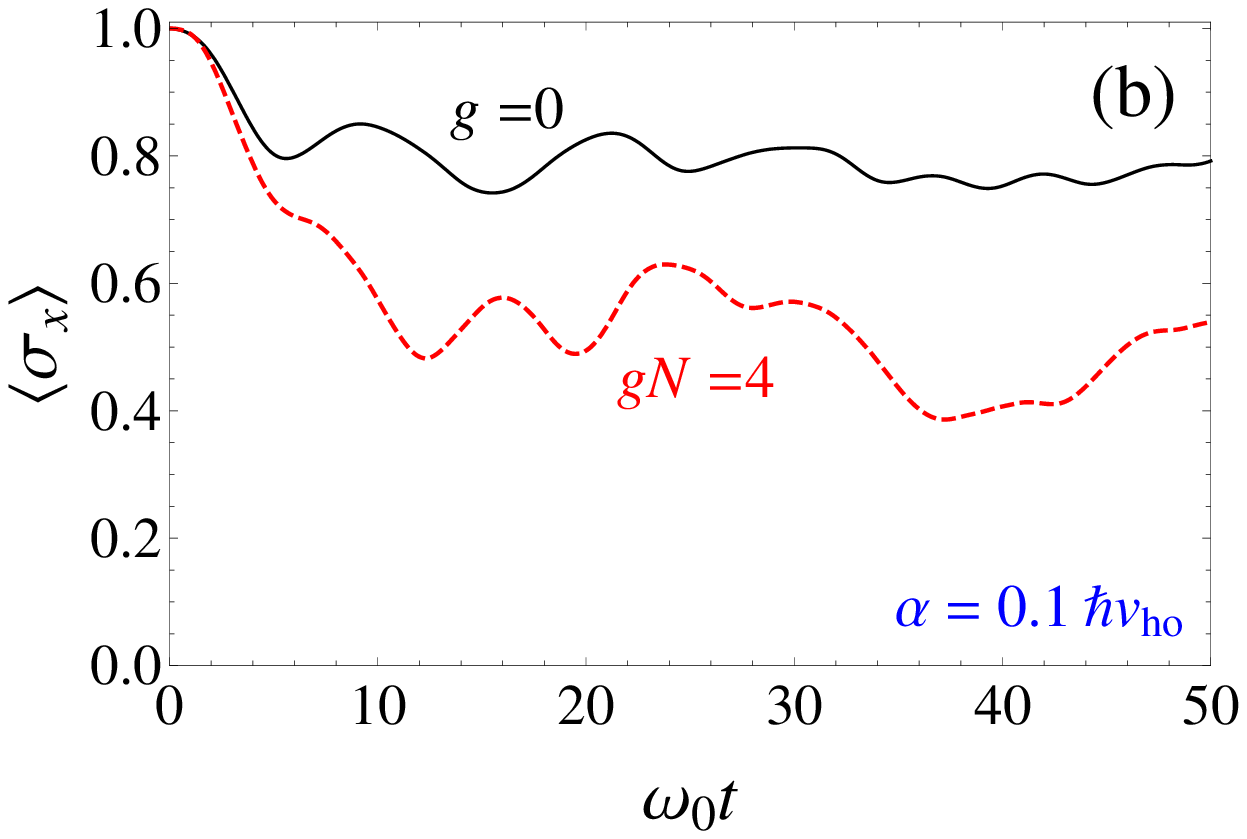}
\end{center}
\caption{Time dependence of the spin components. Solid lines are for noninteracting condensate and dashed lines are for the
interacting (as marked near the plots) one. At small $t$ the behavior of the spin components rather weakly depends 
on the interaction since the 
initial width and position of the wave packet is almost the same for $g=0$ and $gN=4$. With the course 
of time the difference increases. We note that in this 
case $|\langle\sigma_{y}(t)\rangle|\ll 1$  and $\langle\sigma_{x}(t)\rangle\approx P^{1/2}(t)$. 
The qualitative features of the behavior of $\langle\sigma_{y}(t)\rangle,$ dependent 
on the  displacement of the entire condensate, 
do not change  for $gN=4$ since the interaction does not lead to an {\it external} force acting on the condensate.}
\label{figure3}
\end{figure}

\end{widetext}


\begin{thebibliography}{99}

\bibitem{Dyakonov08} \textit{Spin Physics in Semiconductors} Springer Series
in Solid-State Sciences, Ed. by M. I. Dyakonov, Springer (2008).

\bibitem{zhaih2012} H. Zhai, Int. J. Mod. Phys. B \textbf{26}, 1230001
(2012). 

\bibitem{spielman2013} V. Galitski and I. B. Spielman, Nature \textbf{494},
49 (2013). 

\bibitem{EXspielman2011} Y.-J. Lin, K. Jim\'{e}nez-Garc\'{\i}a, and I. B.
Spielman, Nature \textbf{471}, 83 (2011).

\bibitem{EXjin2012} J.-Y. Zhang, S.-C. Ji, Z. Chen, L. Zhang, Z.-D. Du, B.
Yan, G.-S. Pan, B Zhao, Y.-J. Deng, H. Zhai, S. Chen, and J.-W. Pan, Phys.
Rev. Let. \textbf{109}, 115301 (2012).

\bibitem{EXqu2013} Ch. Qu, Ch. Hamner, M. Gong, Ch. Zhang, and P. Engels,
Phys. Rev. A \textbf{88}, 021604 (2013).

\bibitem{EXji2014} S.-C. Ji, J.-Y. Zhang, L. Zhang, Z.-D. Du, W. Zheng,
Y.-J. Deng, H. Zhai, Sh. Chen, and J.-W. Pan, Nature Physics {\bf 10}, 314 (2014).

\bibitem{galitski2008} T. D. Stanescu, B. Anderson, and V. Galitski, Phys.
Rev. A \textbf{78}, 023616 (2008).


\bibitem{stringari2012} Y. Li, G. I. Martone, and S. Stringari, EPL \textbf{%
99}, 56008 (2012); G. I. Martone, Y. Li, L. P. Pitaevskii, and S. Stringari,
Pys. Rev. A \textbf{86}, 063621 (2012). 

\bibitem{zhang2012} Y. Zhang, L. Mao, and Ch. Zhang, Phys. Rev. Lett. 
\textbf{108}, 035302 (2012). 

\bibitem{achilleos2013} V. Achilleos, D. J. Frantzeskakis, P. G. Kevrekidis,
and D. E. Pelinovsky, Phys. Rev. Lett. \textbf{110}, 264101 (2013). 

\bibitem{ozawa2013} T. Ozawa, L. P. Pitaevskii, and S. Stringari, Phys. Rev.
A \textbf{87}, 063610 (2013). 

\bibitem{wilson2013} R. M. Wilson, B. M. Anderson, and Ch. W. Clark, Phys.
Rev. Lett. \textbf{111}, 185303 (2013). 

\bibitem{lu2013} Q.-Q. L\"{u} and D. E. Sheehy, Phys. Rev. A \textbf{88},
043645 (2013). 

\bibitem{brandon2013} B. M. Anderson, I. B. Spielman, and G. Juzeliu\~{n}as,
Phys. Rev. Lett. \textbf{111}, 125301 (2013). 

\bibitem{lindong2014} L. Dong, L. Zhou, B. Wu, B. Ramachandhran, and H. Pu,
Phys. Rev. A \textbf{89}, 011602 (2014). 

\bibitem{Dyakonov} M. I. Dyakonov and V. I. Perel’, Sov. Phys. Solid State {\bf 13}, 3023 (1972).

\bibitem{Adams} E. N. Adams and E. I. Blount, J. Phys. Chem. Solids {\bf 10}, 286 (1959).

\bibitem{Sanchez} L. Sanchez-Palencia, D. Cl\'{e}ment, P. Lugan, P. Bouyer, G. V. Shlyapnikov, and A. Aspect,
Phys. Rev. Lett. \textbf{98}, 210401 (2007).

\bibitem{Flach1} Ch. Skokos, D. O. Krimer, S. Komineas, and S. Flach,
Phys. Rev. E \textbf{79}, 056211 (2009). 

\bibitem{Pikovsky}  A. S. Pikovsky and D. L. Shepelyansky, Phys. Rev. Lett. {\bf 100}, 094101 (2008).

\bibitem{Flach2} S. Flach, D. O. Krimer, and Ch. Skokos,  Phys. Rev. Lett. {\bf 102}, 024101 (2009).

\bibitem{Larcher1} M. Larcher, F. Dalfovo, and M. Modugno, Phys. Rev. A {\bf 80}, 053606 (2009).

\bibitem{Aleiner} I. L. Aleiner, B. L. Altshuler, and G. V. Shlyapnikov,
Nature Physics {\bf 6}, 900 (2010). 

\bibitem{Larcher2} M. Larcher, T. V. Laptyeva, J. D. Bodyfelt, F. Dalfovo, M. Modugno, and S. Flach, 
New J. Phys. {\bf 14}, 103036 (2012).

\bibitem{Min} B. Min, T. Li, M. Rosenkranz, and W. Bao,
Phys. Rev. A {\bf 86}, 053612 (2012). 

\bibitem{Dujardin} J. Dujardin, T. Engl, and P. Schlagheck, cond-mat arXiv:1510.06857.

\bibitem{Zhou}  L. Zhou, H. Pu, and W. Zhang, Phys. Rev. A \textbf{87}, 023625 (2013).

\bibitem{dalfovo1999} F. Dalfovo, S. Giorgini, L. P. Pitaevskii, and S. Stringari, 
Rev. Mod. Phys. {\bf 71}, 463 (1999).

\bibitem{SPbook} L. Pitaevskii and S. Stringari, {\it Bose-Einstein Condensation} (International Series of Monographs on Physics),
Oxford University Press (2003).

\bibitem{gcoupling} For simplicity we assume spin-independent interactions. A possible
spin dependence of $g$ does not have a strong impact on our results. For $^{87}\mathrm{Rb}$ 
in the presence of a transverse harmonic confinement with frequency $\omega _{\perp}=2\pi \times 100$ Hz, the constant ${g}$ 
is of the order of $10^{-3}$ \cite{dalfovo1999,SPbook}. {With $\omega_{0}=2\pi \times 10$ and $N=10^{3}$, the 
mean distance between the atoms in the condensate is of the order of $10^{-4}$ cm. This value greatly 
exceeds the interatomic scattering length of the order of $5\times 10^{-7}$ cm, justifying the applicability 
of the mean-field theory.}

\bibitem{Shklovskii} For the averaging technique, see: A.L. Efros and B.I. Shklovskii, {\it Electronic
Properties of Doped Semiconductors} (Springer, Heidelberg, 1989).

\bibitem{oscillator} For a $^{87}\mathrm{Rb}$ condensate and $\omega _{0}=2\pi \times 10$ Hz, 
we obtain $a_{\rm ho}\approx 3\times 10^{-4}$ cm and $v_{\rm ho}\approx0.02$ cm/s.

\bibitem{normalizedPR} The normalization factor $\sqrt{2\pi}$ is chosen for satisfying condition
$W=\zeta$ for Gaussian wavepackets. 

\bibitem{DP} {The conventional Dyakonov-Perel' mechanism would demonstrate itself in ensembles of spin-orbit coupled Bose-Einstein 
condensates as the total spin dephasing due to different $\langle x(t)\rangle$ and, therefore, $\phi_{12}(t)$ for each condensate in the ensemble. 
The analysis of this mechanism is beyond the scope of our interest.}

\bibitem{Tokatly} Same results can be obtained by gauging spin-orbit
coupling out by a coordinate-dependent spin rotation: 
I. V. Tokatly and E. Ya. Sherman, Phys. Rev. B \textbf{82} 161305 (2010); 
V. A. Slipko and Y. V. Pershin, Phys. Rev. B \textbf{84}, 155306 (2011). 

\bibitem{Supplemental} See Supplemental Material for the shapes of the spinor components 
before and after the fragmentation for a 
stronger spin-orbit coupling and dynamics of spin components in the presence of moderate interatomic interaction. 


\end{thebibliography}
\end{document}